**Fission Decay Widths for Heavy-Ion Fusion-Fission Reactions**

S. G. McCalla[*] and J. P. Lestone

Applied Physics Division, Los Alamos National Laboratory

Los Alamos, NM, 87545, USA

January 29th, 2008

Cross-section and neutron-emission data from heavy-ion fusion-fission reactions are consistent with a Kramers-modified statistical model which takes into account the collective motion of the system about the ground state; the temperature dependence of the location of fission transition points; and the orientation degree of freedom. We see no evidence to suggest that the nuclear viscosity departs from the surface-plus-window dissipation model. The strong increase in the nuclear viscosity above a temperature of ~1 MeV deduced by others is an artifact generated by an inadequate fission model.

The study of the fission of highly excited nuclei remains a topic of great interest [1-5]. For more than twenty years it has been known that the "standard" statistical theory of fission leads to an underestimation of the number of measured pre-scission neutrons emitted in heavy-ion reactions [6-10]. It is generally accepted that the main cause of this discrepancy is associated with the viscosity of hot nuclear matter [11]. Giant dipole resonance γ-ray emission has also been used to infer inadequacies in our models of nuclear fission decay widths [12-15]. These inadequacies have been compensated for by adjusting the viscosity of hot nuclear matter to reproduce experimental data. Few have considered the possibility that the problem with the "standard" model of fission is due to, or partly due to, an incorrect implementation of the model, and a consensus appears to have emerged that strong dissipation sets in rather rapidly above a nuclear temperature of ~1.3 MeV [12].

The Bohr-Wheeler fission decay width is often expressed as [13]

$$\Gamma_f^{BW} = \frac{1}{2\pi \rho_{gs}(E-V_{gs})} \int_0^{E-V_{gs}-B_f} \rho_{sp}(E-V_{gs}-B_f-\varepsilon)\,d\varepsilon, \tag{1}$$

where $\rho_{gs}$ and $\rho_{sp}$ are the level densities at the ground state (stable equilibrium position) and the fission saddle point (unstable equilibrium position), respectively, $E$ is the total excitation energy, $V_{gs}$ is the potential energy at the ground state, and $B_f$ is the fission-barrier height. The level densities are approximated as [13,16]

$$\rho(U) \propto \exp(2\sqrt{aU})/U^2, \tag{2}$$

where $U$ is the thermal excitation energy. The slowing effects of nuclear viscosity are included by using the Kramers-modified [17] Bohr-Wheeler model

$$\Gamma_f = \left(\sqrt{1+\gamma^2} - \gamma\right) \times \Gamma_f^{BW}, \tag{3}$$

---

* Present address, Division of Applied Mathematics, Brown University, RI, 02912, USA

where $\gamma$ is the nuclear viscosity given by $\beta/(2\omega_{sp})$, $\beta$ is the dissipation coefficient, and $\omega_{sp}$ is a measure of the potential curvature at the fission saddle point. Simple arguments show that several pieces of physics are missing from Eqs (1) and (2). These equations contain no terms that allow the fission decay width to change based on the width of the ground-state well. This problem with the statistical model was overcome by Strutinsky [18] who pointed out that the total level density of the system must be calculated taking into account the collective motion about the ground state. This effect increases fission life times by a factor of $T/(\hbar\omega_{gs})$. This issue has been addressed by some [19]. However, many authors continue to ignore this correction.

If the level-density parameter $a$ is a function of deformation then the way Eq. (1) is commonly used becomes invalid at high excitation energy because the locations of the equilibrium points are a function of excitation energy and are defined as the equilibrium points in the level density (or entropy) as a function of deformation, and not as the equilibrium points in the $T=0$ potential energy, $V(q)$. Searching for the equilibrium points in the entropy is the same as searching for the equilibrium points in a temperature-dependent effective potential energy [19,20]

$$V_{eff}(q,T) = V(q) - a(q)T^2. \tag{4}$$

The shape dependence of the level-density parameter can be approximated by the expression [21–23]

$$a(q) \sim c_V A + c_S A^{2/3} B_S(q), \tag{5}$$

where $c_V$ and $c_S$ are constants, and $B_S(q)$ is the ratio of the surface energy relative to that of the spherical system. If the level-density parameter is assumed to be Eq. (5), then the effective potential can be obtained using a $(1-\alpha T^2)$ dependence of the surface energy, where

$$\alpha = c_S A^{2/3}/E_S^0 = c_S \times 0.059 \text{MeV}^{-1} \text{ for } A \sim 200 \tag{6}$$

and $E_S^o$ is the surface energy of the spherical system. For a particular model, Töke and Swiatecki [23] obtain $c_S \sim 0.27$ MeV$^{-1}$. This gives an estimate for the value of $\alpha \sim 0.016$ MeV$^{-2}$. $c_S$ is known to be very sensitive to the assumed properties of nuclear matter and to other approximations [24]. Other estimates of $c_S$ [21-26] give values of $\alpha$ that range from 0.007 to 0.022 MeV$^{-2}$. Initially, we shall assume $\alpha$=0.016 MeV$^{-2}$.

The locations of fission transition points do not change much up to a temperature of ~1 MeV. However, there is a dramatic change in the locations of the transition points above $T$~1 MeV (see Fig. 1). The dashed vertical lines in Fig. 1 show that the equilibrium points in the effective potential correspond to equilibrium points in the entropy. If the transition point is incorrectly assumed to equal the $T$=0 value (independent of temperature) then the entropy at the transition point will be increasingly overestimated with increasing temperature. This causes the mean fission life time to be increasingly underestimated as the temperature increases above ~1 MeV.

The statistical model of the fission of rotating systems must determine the total level density and the number of fission transition states, taking into account the level density associated with both the shape and orientation degrees of freedom. In the limit of a small deviation from the spherical shape, the effective moment of inertia, $I_{eff}(q)$, is large and the rotational energy becomes independent of the orientation of the symmetry axis relative to the total spin. In this case, including the orientation degree of freedom increases the level density by $2J$+1 [28]. For an arbitrary deformation, this multiplication factor associated with the orientation degree of freedom [28] is less than $2J$+1,

$$f = \sum_{K=-J}^{J} \exp\left(-K^2 \hbar^2 / (2T I_{eff}(q))\right), \quad (7)$$

where $K$ is the spin about the fission axis. This decrease in the level density with increasing deformation slows fission at high spins relative to widths obtained using Eqs (1) and (2).

Including these three effects, the statistical model fission decay width for a rotating system, using the number of transition states and the total level density given by [18,28]

$$N_{TS} = \sum_K \int \rho_{sp}(E - V_{sp}(K,T) - \varepsilon) d\varepsilon, \text{ and} \quad (8)$$

$$\rho = \sum_K \frac{T_{gs}}{\hbar \omega_{gs}(K,T)} \rho_{gs}(E - V_{gs}(K,T)), \quad (9)$$

is [29]

$$\Gamma_f = \sum_K P(K)\Gamma_f(K) / \sum_K P(K), \quad (10)$$

where $P(K)$ is the probability that the system is in a given $K$ state,

$$P(K) \propto \frac{T_{gs}(K)}{\hbar \omega_{gs}(K,T)} \rho_{gs}(E - V_{gs}(K,T)). \quad (11)$$

$\Gamma_f(K)$ is the fission decay width if the system were restricted to a single $K$. To include the effect of the nuclear viscosity, the fission decay width as a function of $K$ should be determined including the Kramers' reduction factor using

$$\Gamma_f(K) = \left(\sqrt{1+\gamma^2} - \gamma\right) \times \frac{\hbar \omega_{gs}}{2\pi} \exp(-\frac{2B_f}{T_{gs} + T_{sp}}), \quad (12)$$

where $\omega_{sp}$, $\omega_{gs}$ and $B_f$ are all functions of temperature and $K$, and determined using an effective potential with a $(1-\alpha T^2)$ dependence of the surface energy. $T_{gs}$ and $T_{sp}$ are the temperatures at the equilibrium points. The statistical model code JOANNE4 [29] includes the three effects discussed here.

Calculated cross sections and pre-scission emission are very sensitive to the assumed $T=0$ potential-energy surface and the assumed deformation dependence of the level-density parameter, but depend only linearly on the nuclear viscosity. Therefore, extreme caution must be applied if attempting to infer the viscosity from cross-section and emission data. We believe the nuclear dissipation has been well constrained by the surface-plus-window dissipation model [30,31] using the mean kinetic energy of fission fragments and the widths of isoscalar giant resonances. For the range of typical fission saddle-point deformations encountered in heavy-ion fusion-fission reactions with compound nuclei mass numbers from $A_{CN}$ ~170 to 220, the dissipation coefficient in this model is within 10% of $3\times10^{21}$ s$^{-1}$ and independent of temperature.

To confirm the validity of the Kramers-modified statistical model we compare results obtained using Eqs (10)-(12) to dynamical calculations. We assume the shape degree of freedom is governed by a Langevin equation [32] and couple the orientation degree of freedom ($K$-states) with the heat bath [29]. We assume the same temperature dependence effective potential $V_{eff}(q,T)$, the same dissipation coefficient, and the same inertia [33] for both our statistical and Langevin calculations. Langevin calculations performed by others do not include a coupling between the orientation degree of freedom and the heat bath, and therefore, do not allow the $K$ states to equilibrate. The Langevin calculations of others underestimate the fission lifetime because only the $K=0$ fission barrier is sampled, instead of an equilibrated distribution containing higher $K\neq0$ barriers. Fig. 2 shows estimates of the mean fission life time of $^{210}$Po systems formed by the reaction $^{18}$O + $^{192}$Os, as a function of the initial excitation energy. The relationship between excitation energy and average spin of the fissioning systems is determined using measured fusion and evaporation-residue cross sections [34]. The solid curve shows Kramers-modified statistical model calculations using Eqs (10) and (11), with the fission decay width as a function of $K$ determined using Eq. (12) with $\omega_{sp}$, $\omega_{gs}$ and $B_f$ obtained using the effective potential as a function of both $T$ and $K$, as performed by JOANNE4 [29]. Model parameters are $a=A/8.6$ MeV$^{-1}$, $\beta=3\times10^{21}$ s$^{-1}$, and $\alpha=0.016$ MeV$^{-2}$. These calculations are consistent with the corresponding two-dimensional (shape and orientation) Langevin calculations shown by the solid circles in Fig. 2. In the statistical model the dissipation influences the fission life time through the Kramers' factor. In the Langevin calculation the dissipation controls the size of random kicks in the collective motion and the rate these kicks are converted back into heat. The agreement between these two methods confirms that if the Kramers-modified statistical model is implemented correctly, then the results are in agreement with dynamical calculations.

Dioszegi et al. [13] assume $a_{sp}/a_{gs} \sim 1.04$ when inferring the nuclear viscosity of hot rotating $^{224}$Th nuclei. The dashed curve in Fig. 2 shows estimates of the fission life-time of $^{210}$Po obtained using the standard Bohr-Wheeler model with $a_{sp}/a_{gs}=1.04$. These calculations are a factor of two lower than the more complete calculations shown by the solid curve and circles at $E_i \sim 40$ MeV, and more than a factor of 20 low at $E_i \sim 90$ MeV (see Fig. 2). The solid curve in Fig. 3 shows the nuclear viscosity as a function of excitation energy needed to force the standard Bohr-Wheeler model with $a_{sp}/a_{gs}=1.04$ to be in agreement with the calculations shown by the solid curve in Fig. 2. This artificial excitation-energy dependence of the nuclear viscosity is similar to the excitation-energy dependence deduced by Dioszegi et al. [13]. This result suggests that the strong excitation-energy dependence of the nuclear viscosity [13] and the rapid onset of the dissipation at nuclear excitation energies above ~40 MeV [12] are artifacts generated by an incomplete model of the fission process.

To avoid complexities associated with sensitivities to assumed shell corrections for very fissile systems and assumed fusion spin distributions for light systems, we restrict our use of the statistical model code JOANNE4 to reactions with compound nucleus masses $A_{CN} \sim 170$ to 220. We model the fusion process and adjust the nucleus-nucleus potential and shape of the target nucleus to obtain fits to measured fusion excitation functions [10,34]. The corresponding calculated fusion spin distributions are used as input into the statistical model calculations. All JOANNE4 calculations presented here assume $a=A/8.6$ MeV$^{-1}$ and $\beta=3\times10^{21}$ s$^{-1}$. The only free parameters are $\alpha$ and a scaling of the MLDM radii from their default values. A scaling of $r_S=1.0$ gives the default MLDM with fission barriers in agreement with the finite-range liquid-drop model (FRLDM) [35]. If the nuclear radii are made larger by increasing $r_S$ above 1, then the surface energy increases and the Coulomb energy decreases. This makes the systems more stable and increases the fission barriers. For each reaction considered here, the free parameters $\alpha$ and $r_S$ are adjusted to reproduce a single fission cross section and a single pre-scission neutron multiplicity at the same projectile kinetic energy, corresponding to the second lowest pre-scission neutron multiplicity measurement. Fig. 4 shows how the $E_{lab}$~103 MeV $^{18}$O + $^{192}$Os data constrain the adjustable parameters to $\alpha$=0.017±0.006 MeV$^{-2}$ and $r_S$=1.002±0.002. Fig. 5 shows the parameters $\alpha$ and $r_S$ for five reactions. Fig. 6 shows the model predictions for the projectile energy dependence of fission and residue cross sections and pre-scission neutron multiplicities, using the $\alpha$ and $r_S$ values represented by the circles in Fig. 5. These predictions are consistent with the data. To reproduce this data set, the model calculations of others would require either large fission dynamical delays [10] or a strong temperature dependence of the nuclear viscosity as shown in Fig. 3.

Other authors have assumed that their ability to model nuclear fission is complete enough that the properties of nuclear dissipation can be extracted from cross section and pre-scission emission data. We assume that the nuclear dissipation near fission transition points has been previously constrained to be $\beta$~3×10$^{21}$ s$^{-1}$ [30,31]. Fusion-fission cross-section and pre-scission neutron data for $A_{CN}$ = 170–220 are consistent with a Kramers-modified statistical model of fission, the FRLDM [35], and a shape dependence of the level-density parameter in the range of theoretical estimates [21-26].

LA-UR-08-0207

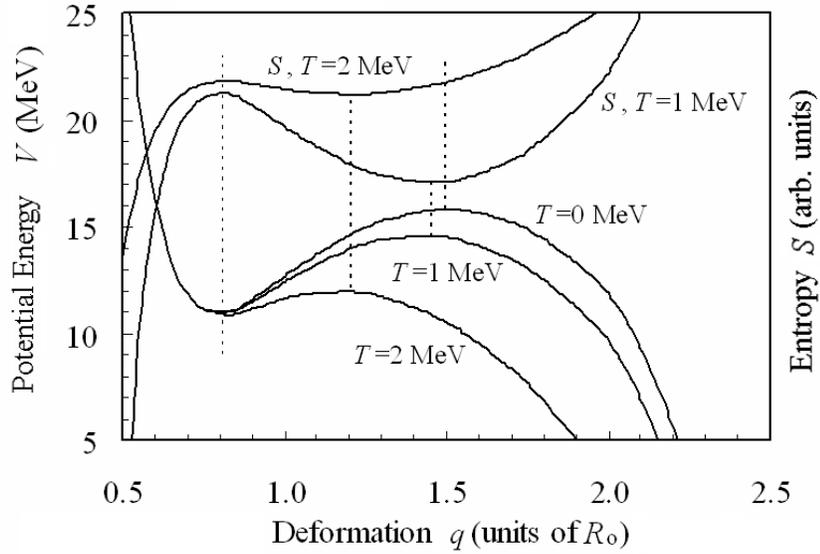

Fig. 1. The modified liquid-drop model (MLDM) [27] potential energy, $V(q)$, as a function of deformation for $^{210}$Po with $J=50$, along with the corresponding effective potential energies, $V_{eff}(q,T)$, at $T=1$ and 2 MeV assuming $\alpha=0.016$ MeV$^{-2}$. Also shown is the deformation dependence of the corresponding entropies, $S(q,E)$. The dashed vertical lines are to guide the eye.

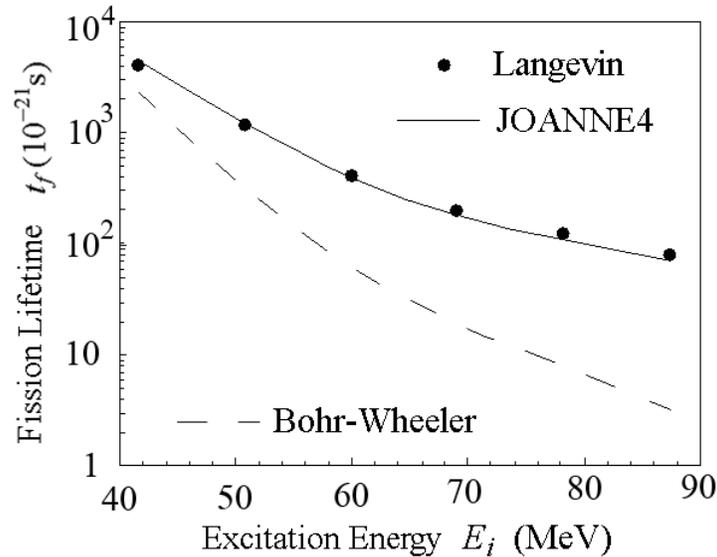

Fig. 2. Various estimates of the mean fission life time of $^{210}$Po systems formed by the reaction $^{18}$O + $^{192}$Os, as a function of the initial excitation energy (see text).

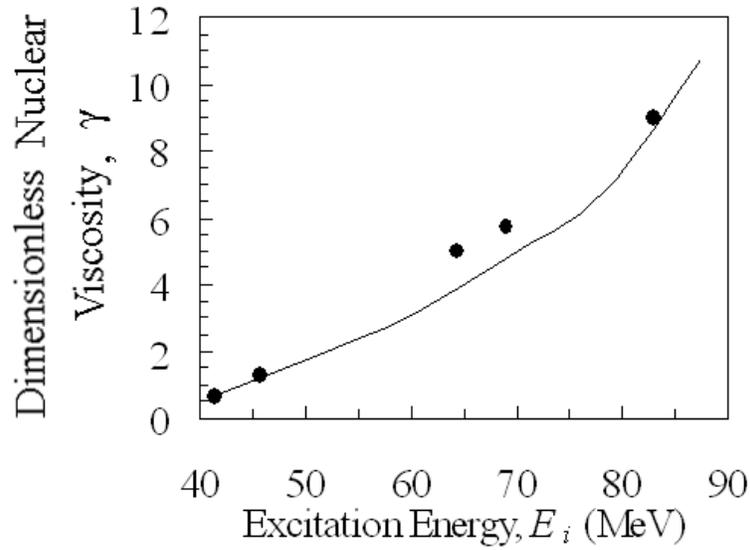

Fig. 3. The solid line shows the nuclear viscosity as a function of excitation energy needed to force the Kramers-modified standard Bohr-Wheeler model with $a_{sp}/a_{gs}=1.04$ to be in agreement with the calculations shown by the solid curve in Fig. 2. The symbols show the nuclear viscosity inferred by ref [13].

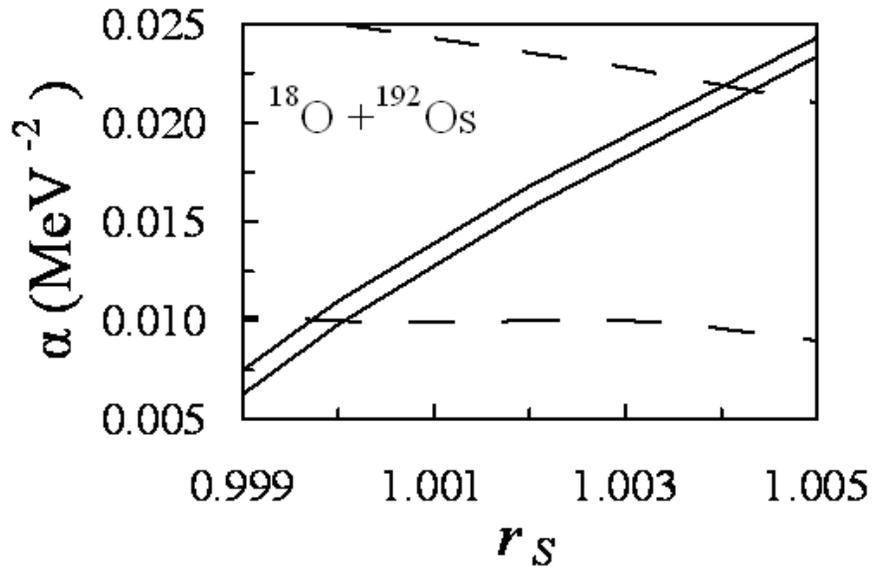

Fig. 4. The $E_{lab} \sim 103$ MeV $^{18}$O + $^{192}$Os fission cross section [34] and neutron multiplicity [10] constrain the parameters $\alpha$ and $r_S$ to the regions between the solid and dashed curves, respectively.

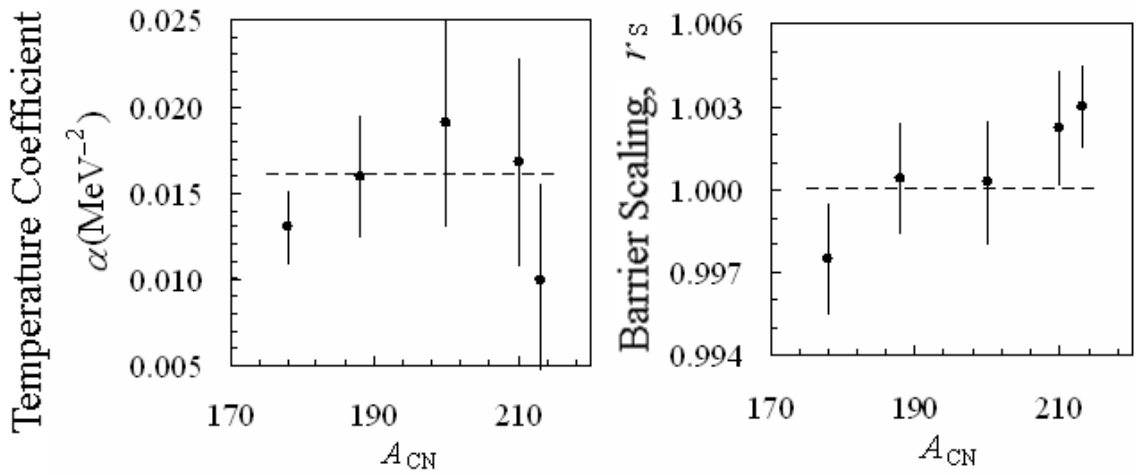

Fig. 5. Fit parameters α and $r_S$ for five reactions. The dashed lines show the values corresponding to the model calculations of refs [23] and [35].

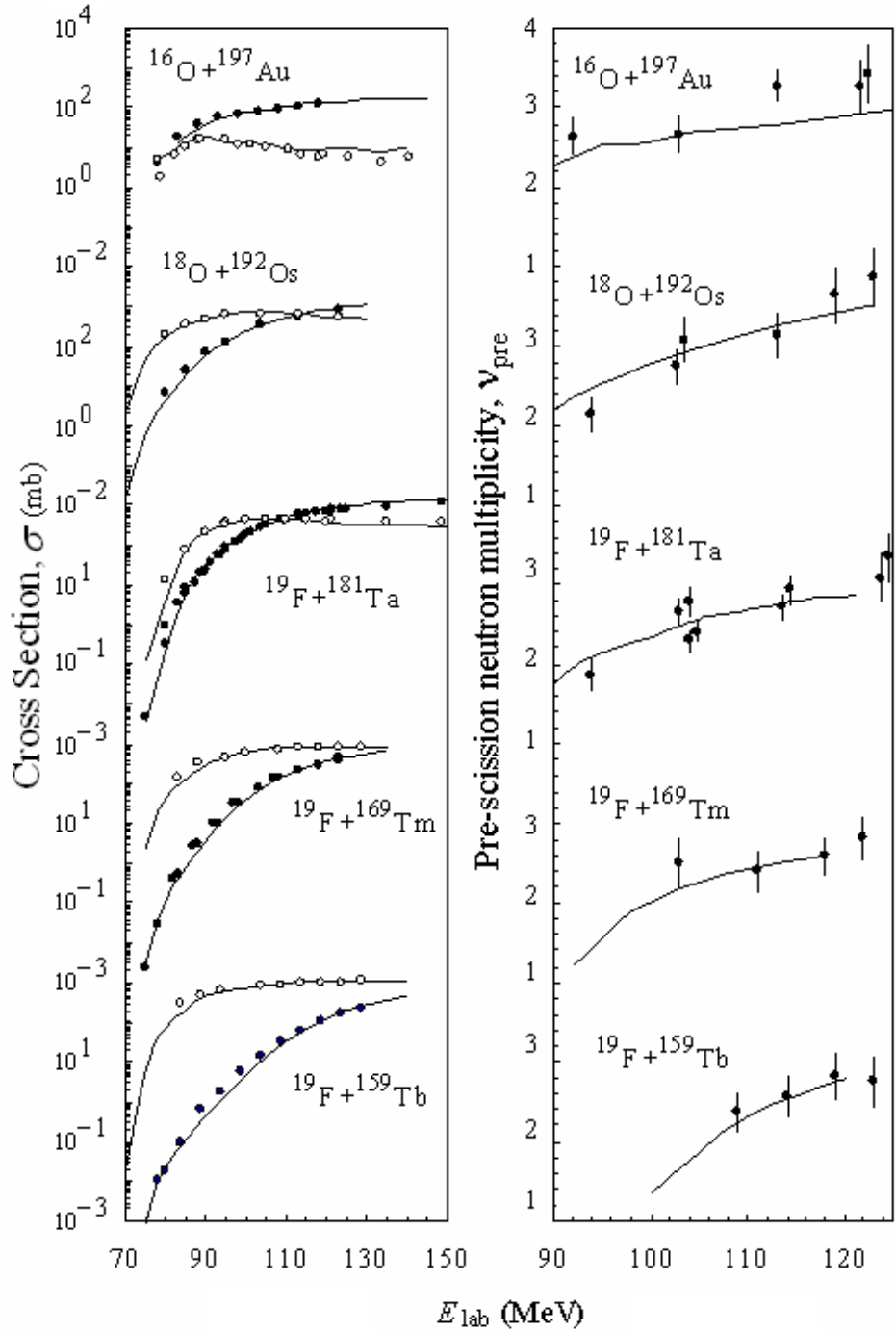

Fig. 6. JOANNE4 model predictions for the projectile energy dependence of cross sections and pre-scission neutron multiplicities for five reactions. The experimental data are from refs [10,34,36-39]. The fission and residue cross sections are shown by solid and open symbols, respectively.

LA-UR-08-0207